\documentclass[a4paper,fleqn,usenatbib]{mnras}

\usepackage{times}

\usepackage[T1]{fontenc}
\usepackage{ae,aecompl}

\usepackage{graphicx}	
\usepackage{amsmath}	
\usepackage{amssymb}	
\usepackage{epsfig}
\usepackage{pdflscape}

\def\msun{\hbox{M$_\odot$}}

\def\t4{\hbox{t$_{\rm 4}$}}

\def\cm3{\hbox{cm$^{-3}$}}

\input psfig.sty
\input epsf.sty

\title[Multiple populations in Kron 3]{Kron 3: a fourth intermediate age cluster in the SMC with evidence of multiple populations\thanks{Based on observations made with ESO telescopes at the La Silla Paranal Observatory under Programme ID 096.B-0618(B).}}

\author[Hollyhead et al.]{
K. Hollyhead$^{1}$\thanks{E-mail: kathie.hollyhead@astro.su.se)},
C. Lardo$^{2}$,
N. Kacharov$^{3}$,
N. Bastian$^{4}$,
M. Hilker$^{5}$,
M. Rejkuba$^{5,6}$,
\newauthor A. Koch$^{7,8}$,
E. K. Grebel$^{8}$ and
I. Georgiev$^{3}$
\\
$^{1}$ Department of Astronomy, Oscar Klein Centre, Stockholm University, AlbaNova, Stockholm SE-106 91, Sweden\\
$^{2}$Laboratoire d'astrophysique, Ecole Polytechnique F\`ed\`erale de Lausanne (EPFL), Observatoire de Sauverny, CH-1290 Versoix, Switzerland\\
$^{3}$Max-Planck-Institut f\"{u}r Astronomie, K\"{o}nigstuhl 17, D-69117 Heidelberg, Germany\\
$^{4}$Astrophysics Research Institute, Liverpool John Moores University, 146 Brownlow Hill, Liverpool L3 5RF, UK\\
$^{5}$European Southern Observatory, Karl-Schwarzschild-Stra\ss{}e 2, D-85748 Garching bei M\"{u}nchen, Germany\\
$^{6}$Excellence Cluster Universe, Boltzmannstr. 2, 85748, Garching, Germany\\
$^{7}$Department of Physics, Lancaster University, Lancaster LA1 4YB, UK\\
$^{8}$Astronomisches Rechen-Institut, Zentrum f\"{u}r Astronomie der Universit\"{a}t Heidelberg, M\"{o}nchhofstr. 12-14, 69120 Heidelberg, Germany\\
}

\date{Accepted XXX. Received YYY; in original form ZZZ}

\pubyear{2016}

\begin{document}
\label{firstpage}
\pagerange{\pageref{firstpage}--\pageref{lastpage}}
\maketitle

\begin{abstract}
We present the results of a spectroscopic study of the intermediate age ($\approx 6.5$ Gyr) massive cluster Kron 3 in the Small Magellanic Cloud. We measure CN and CH band strengths (at $\simeq 3839$ and $4300 $ \AA~respectively) using VLT FORS2 spectra of 16 cluster members and find a sub-population of 5 stars enriched in nitrogen. We conclude that this is evidence for multiple populations in Kron 3, the fourth intermediate age cluster, after Lindsay 1, NGC 416 and NGC 339 (ages 6-8 Gyr), to display this phenomenon originally thought to be a unique characteristic of old globular clusters. At $\approx6.5$ Gyr this is one of the youngest clusters with multiple populations, indicating that the mechanism responsible for their onset must operate until a redshift of at least 0.75, much later than the peak of globular cluster formation at redshift $\sim 3$. 
\end{abstract}




\section{Introduction}
\label{sec:intro}


Multiple populations (MPs) have been found in old, globular clusters (GCs, $>10$ Gyr) in both the Milky Way \citep[e.g.;][]{gratton12} and the LMC \citep{mucciarelli09, mateluna12} as well as the SMC \citep{dalessandro16} and Fornax dwarf galaxy \citep{larsen14}.They are characterised spectroscopically by light element abundance variations and anticorrelations \citep[e.g. O-Na anticorrelation;][]{carretta09} and photometrically by spreads or splits in the main sequence and/or red giant branches in certain filters \citep[e.g.][]{marino08, piotto15}. 

The presence and mechanism behind the onset of MPs has significant consequences for globular cluster formation theories \citep[e.g.;][]{decressin07,dercole08,demink09,bastian13}. There are currently several proposed mechanisms for the formation of MPs, though all have difficulties recreating the required light element abundance variations that are observed across all GCs \citep[][]{bastian15b,renzini15} and also run into other significant issues, the most prominent being the mass-budget problem \citep[e.g.][]{larsen12,bastianlardo15,kruijssen15}. Additionally, sufficient gas reservoirs that would be required for the formation of a second generation at the ages required by the current GC formation theories have also not been found in young massive clusters \citep{ivan15, longmore15}. Having a working theory for the formation and evolution of globular clusters is important for overall galactic formation theories, particularly as populations of stars with similar properties have also been identified within the bulge of the Milky Way \citep{schiavon17}.

Mass has been found to be an important factor in whether MPs form in clusters \citep{carretta10, schiavon13, milone17}. The lowest mass Milky Way GC with MPs discovered recently is NGC 6535 at $10^{3.58}$\msun~\citep{milone17,bragaglia17}, or potentially ESO452-SC11 at 6.8$\pm$3.4$\times$10$^3$\msun~\citep{simpson17}. However, massive clusters in the age range of 1-2 Gyr in the Magellanic Clouds of masses equal or greater than those of GCs have also been found to show a lack of evidence for MPs \citep[e.g. NGC1806 and NGC419;][]{mucciarelli14, martocchia17}, indicating that age also plays an important role. Intriguingly, the phenomenon of extended main sequence turn-offs (eMSTOs) has been identified in clusters younger than 2 Gyr \citep[e.g.][]{mackey08}, which were originally attributed to age spreads of $\sim$ 200-700 Myr \citep[e.g.][]{goudfrooij14}. However the magnitude of the spread was found to be proportional to the age of the cluster, and therefore unlikely to be real \citep{niederhofer15, milone15}. It is currently thought that stellar rotation can explain this observation \citep[e.g.][]{bastian09, niederhofer15}. Open clusters of comparable age (6-9 Gyr) and mass ($\sim10^4$\msun) to GCs such as NGC 6791 or Berkeley 39 have also been found to lack MPs \citep{bragaglia12,bragaglia14,cunha15}. 

Until recently, there had existed a gap in the age ranges of clusters studied with the aim of looking for MPs, from $\approx 2-10$ Gyr. It was evident that observing massive clusters in this age range could help constrain the age at which MPs form and potentially the mechanism by which they are created. Lindsay 1, at $\sim8$ Gyr old \citep{mighell98, glatt08}, was the first non-traditional ancient globular cluster to show evidence for MPs in the form of a statistically significant nitrogen spread compared to a negligible spread in carbon \citep{hollyhead17}. Six stars out of the 16 member stars were nitrogen enriched and belonged to a secondary sub-population within the cluster. This result was confirmed with HST photometry and a split in the RGB \citep{niederhofer17}. Overlapping stars in the catalogues of the two studies indicated that stars with enhanced nitrogen lay on the secondary RGB when the cluster is imaged in a combination of {\em special filters} sensitive to N variations. In addition to Lindsay 1, NGC 416 and NGC 339 were then also found to host MPs from photometry \citep{niederhofer17}.

In this paper we present the results for a further intermediate age, massive, metal-poor \citep[3.9-5.84$\times10^5$\msun, \text{[Fe/H] =} $-1.08$ dex;][]{glatt11} cluster in the SMC, Kron 3 (hereafter K3). At 6.5 Gyr old \citep{glatt08} this is one of the youngest clusters with a detailed spectroscopic search for MPs, along with NGC 416 and NGC 339, both at $\sim 6$ Gyr \citep{glatt08}. In order to look for MPs we carried out a similar study to that of Lindsay 1 in \cite{hollyhead17} (hereafter H17) and obtained low resolution FORS2 spectra of 35 targets in the direction of the cluster. Using this data we calculate CN and CH band strengths \citep[e.g.][]{kayser08,pancino10, lardo12} and use stellar sythesis to find [C/Fe] and [N/Fe] to look for MPs. 

In \S~\ref{sec:data} we discuss our data taken using FORS2 on the VLT describing briefly how the spectra were reduced. \S~\ref{sec:members} outlines our extensive and strict membership criteria for determining which of our stars belonged to K3 and we then outline our method for calculating CN and CH band strengths and abundances in \S~\ref{sec:indices}. Finally, \S~\ref{sec:results} shows our results for this cluster and in \S~\ref{sec:concs} we discuss our conclusions.

\section{Observations and data reduction}
\label{sec:data}

Our data for K3 was taken during the same observing run as Lindsay 1, program ID 096.B-0618(B), P.I. N. Kacharov, as described in H17. Using multi-object spectroscopy with FORS2 on the VLT at the Paranal observatory in Chile, we obtained six science images centred on K3, along with bias frames and flat fields. It was necessary to use the GRIS 600B+22 grism to observe CN and CH bands at 3839~\AA{} and 4300~\AA{} respectively. Archival pre-imaging was available for photometry (ESO-programme: 082.B-0505(A) P.I. D. Geisler) in the V and I bands, for which we have errors of $sim 0.05$. The resolution of the spectra is $R = \lambda/\Delta \lambda \simeq 800$, covering a nominal spectral range of $\sim 3300 - 6600$ \AA{}. The sampled spectral range varies from star to star and depends on their position on the instrument field of view.

Similarly as done for L1, we centered the master chip on the centre of the cluster, with the slave chip at the southern edge. 35 targets were also chosen for K3, with primary targets selected from the CMD sampling the lower RGB to avoid abundance contamination due to stellar evolution and the first dredge-up \citep{cohen02,kayser08,pancino10}. The centre of the cluster could not be sampled due to crowding. 


Using \textsc{iraf}, we reduced the spectra by subtracting bias frames, applying flat fields, and wavelength calibrating and extracting the 1D spectra. Cosmic rays were removed with the L.A.Cosmic \textsc{iraf} routine \citep{lacos}. During extraction, the apertures showed little curvature, similar to L1 and so we allowed the \textit{apall} task to account for this without applying an additional correction.

\begin{figure}
	\includegraphics[width=8.5cm]{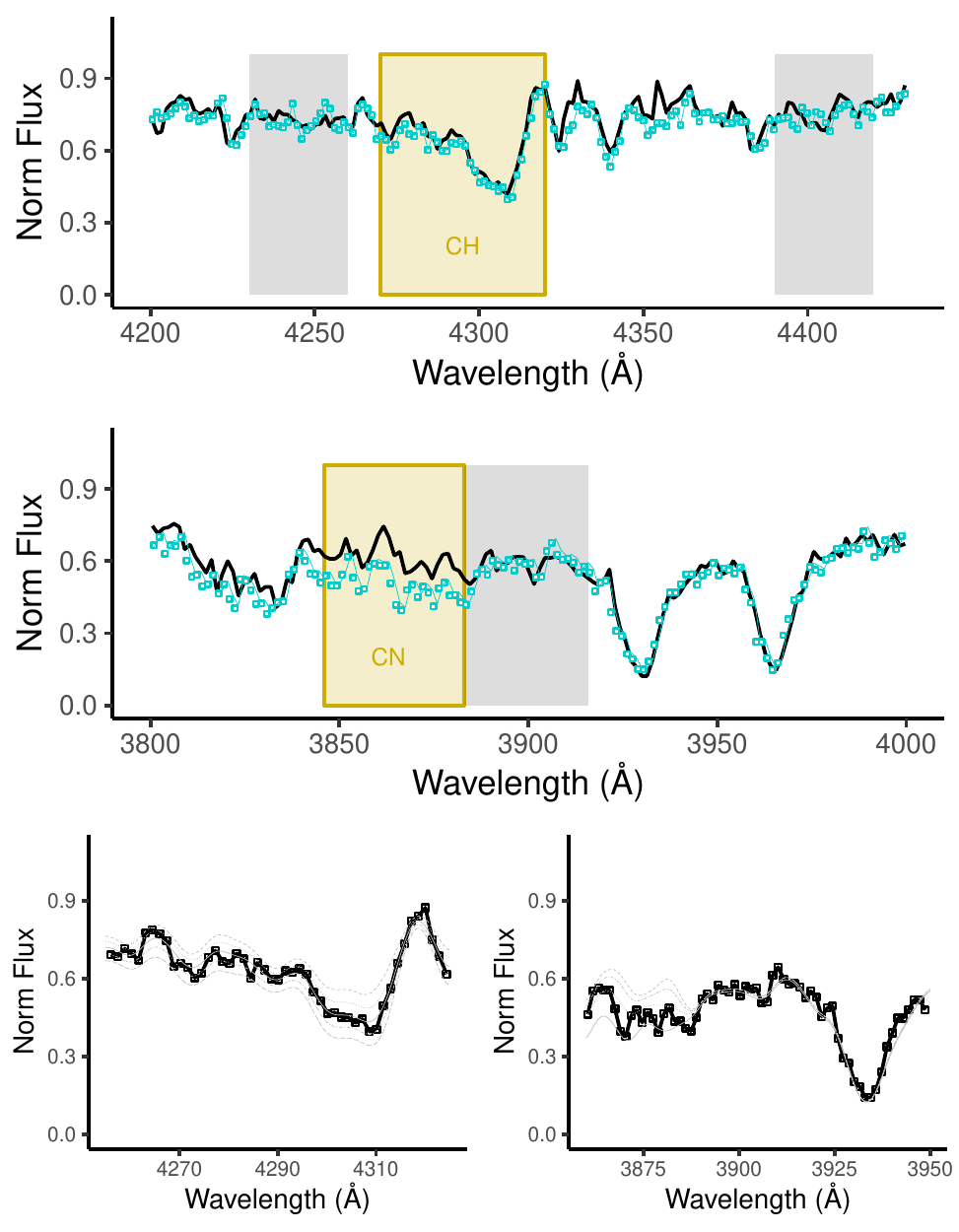}
	\caption{Example spectra of two member stars in K3 (stars 225 and 3910). The stars have similar Teff and log(g) to illustrate the difference seen in the CN band at 3839~\AA{} and the similarity in the CH band at 4300~\AA{}. 
	The yellow regions represent the spectral regions from which we measured the CN and CH indices. The grey areas are the respective continuum windows. The bottom left panel shows observed (black) and synthetic (grey) spectra around CH band for the star 3910. The grey lines are the syntheses computed with C abundance altered by 
--0.8, --0.6, --0.4, --0.2, +0.0 dex (from bottom to top). The bottom right panel shows the same as the left hand panel but for the CN feature. The synthetic spectra show the syntheses computed with N abundance altered by --0.8, --0.4, +0.0, +0.6 dex (from bottom to top). }
	\label{fig:spectra}
\end{figure}

Fig.~\ref{fig:spectra} shows example spectra of two member stars of K3. The bands adopted to measure S$\lambda3839$ and CH$\lambda$4300, and used for the spectral synthesis to measure [C/Fe] and [N/Fe] are highlighted on the spectra. We have chosen two stars with very similar effective temperatures and log(g) to show the differences in the CN band, which traces nitrogen.

\begin{figure}
\centering
	\includegraphics[width=0.7\columnwidth]{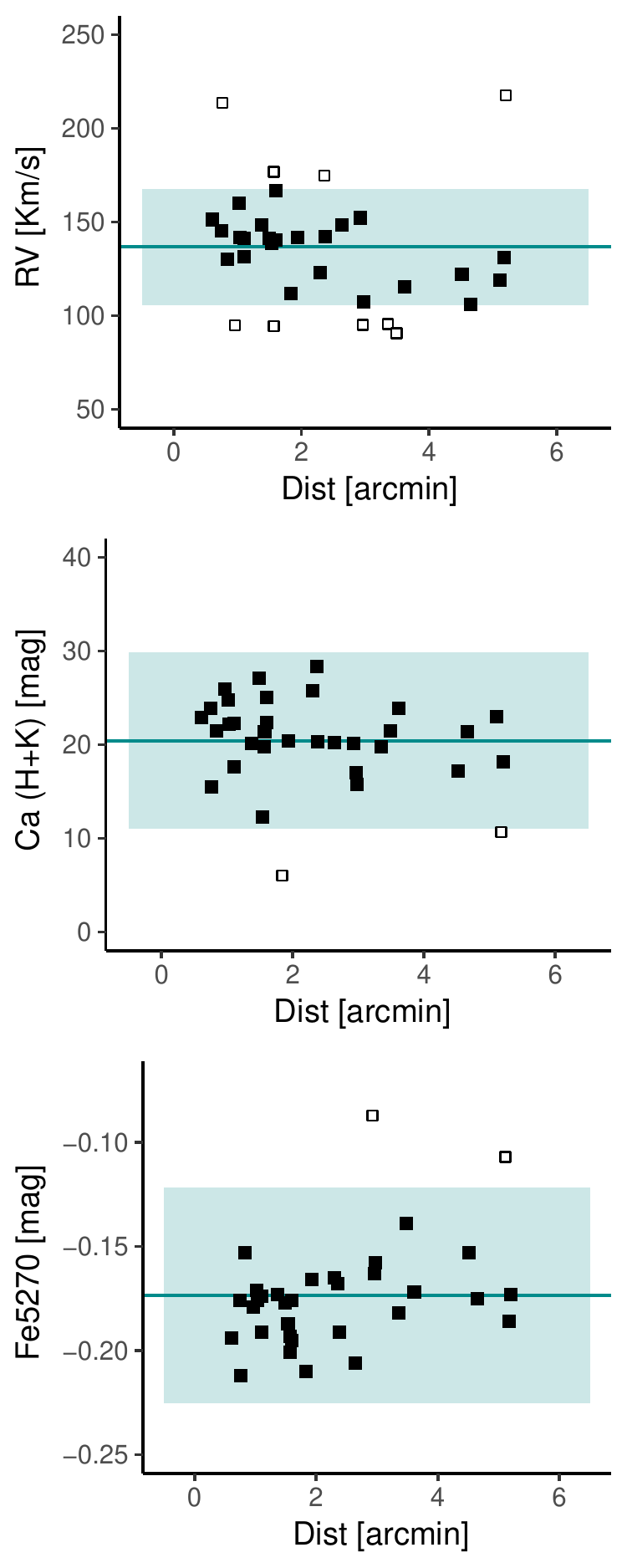}
	
	\caption{The radial velocity, Ca{\sc ii} (H+K) and Fe5270 measurements for all stars against their distances from the centre of the cluster in arcminutes. 
	Filled symbols are considered members and empty ones non-members. 9 stars were removed due to outlying radial velocities, two to Ca{\sc ii} (H+K) and two from Fe5270, though several of these overlap.}
	
	\label{fig:rvhkfe}
\end{figure}

\begin{figure}
\centering
	\includegraphics[width=0.9\columnwidth]{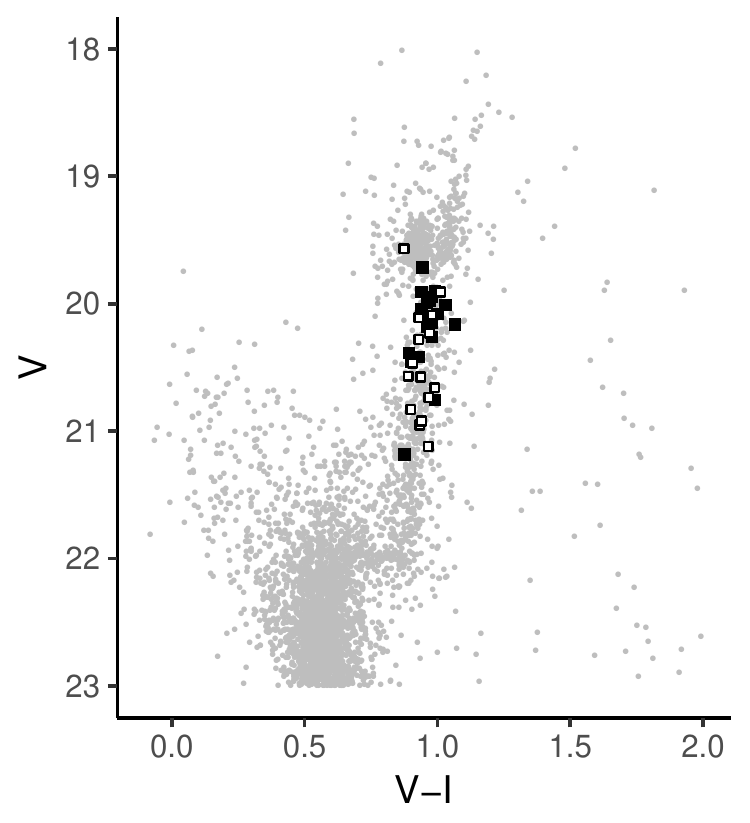}
	\caption{One of our criteria used to determine cluster membership. The plot shows the CMD for all stars in our analysis with stars which survived our selection based on radial velocity, Ca{\sc ii} (H+K) and Fe5270 measurements as filled symbols. Empty symbols are rejected stars. All stars appear to lie on the RGB, however the star at $V \sim 19.7$ was removed due to its high luminosity and the possibility of it being affected by mixing.}
	\label{fig:rvplot}
\end{figure}

\section{Cluster membership}
\label{sec:members}

Due to the low resolution of the spectra, it is more difficult to constrain cluster membership solely via radial velocity measurements. Therefore, we have employed a number of criteria to more accurately determine cluster membership as described in detail in H17. Fig.~\ref{fig:rvhkfe} and Fig.~\ref{fig:rvplot} show how we used our criteria to determine which stars are true cluster members.

Fig.\ref{fig:rvhkfe} shows the radial velocities along with Ca{\sc ii} (H+K) and Fe5270 band strength measurements (see \S~\ref{sec:cnch}) for all stars, plotted against their distances from the centre of the cluster in arcminutes. The first panel shows our radial velocities measured in \textit{fxcor} in \textsc{iraf} adopting a likely member star as a template with its radial velocity measured first with the \textit{rvidlines} routine. The teal line indicates the previously observed velocity for K3 at 135.1 km/s by \citet{parisi15}. We decided on cluster members using a 1 $\sigma$ ($\approx 30$ km/s, which is also compatible with the precision of FORS2 with our setup) allowance from this value, indicated by the blue background. Filled symbols are members, while empty squares are outside of our range and considered non-members. Nine stars were removed using radial velocities.


The second panel shows Ca{\sc ii} (H+K) index measurements for all stars. In this case we used a 2$\sigma$ ($\approx 9$ mag) cut off from the median for member stars. Two stars were identified as non-members from this index. Finally, the lowest panel shows the Fe5270 band strengths. Again a 2$\sigma$ ($\approx 0.05$ mag) cut was used and two stars were identified. Many of the criteria had overlapping stars identified as non-members. The spread in these measurements is larger than we found for L1, where we were able to use a cut-off of 1$\sigma$ from the median of both Fe5270 and Ca{\sc ii} (H+K). With K3, however, using 1$\sigma$ would remove potential cluster members and many more than those outside of the acceptable radial velocity range. Therefore we use 2$\sigma$ as our range for members. 

Fig.\ref{fig:rvplot} shows the CMD for all stars in the photometry of K3, with stars which survived our previous selection showed as filled symbols. All stars lie on the RGB, however we removed the bright star 791 at $V \sim 19.7$ as it could be more strongly affected by evolutionary mixing due to its luminosity \citep{gratton00, angelou15, dotter17}.

Finally, spectra with S/N$\leq$10  (per pixel) in the CN band region were rejected. Spectra with significant defects (spikes, holes) in the measurement windows were also rejected.
These cuts left 16 member stars for K3 across both chips (see Table~\ref{tab:data}).

\section{Index measurement and spectral synthesis}
\label{sec:indices}

\subsection{CN and CH band strengths}
\label{sec:cnch}

As per H17 we calculated the $UV$ CN and G CH band strengths (S($\lambda 3839$) and  CH($\lambda$4300) respectively) using the definitions from \citet{Norris81, Worthey94} and \citet{Lardo13} with error measurements estimated as per \citet{Vollmann06}. The average S/N for the CN and CH bands is $\sim25$ and $\sim40$, respectively.
As discussed in e.g. \citet{lardo12}, the CH band at 4300\AA~ is not affected by the change in spectral slope from atmosphere or instrumental effects thanks to two continuum bandpasses. On the other hand, one has to rely only on a single continuum bandpass in the red part of the spectral feature for the 3883\AA~CN band (see Fig.~\ref{fig:spectra}). Following \citet{cohen02,cohen05}, we normalised the stellar continuum in the spectrum of each star and then measured the absorption within both the CH and CN bandpasses.  The polynomial fitting used a 6$\sigma$ high and 3$\sigma$ low clipping, running over a five pixel average. 
By fitting the continuum, we were able to directly compare the indices measured in this section and the abundances derived from spectral synthesis in Sect.~\ref{sec:candn}.

Fe5270 and Ca{\sc ii} (H+K) used previously to constrain membership in \S~\ref{sec:members} were calculated in a similar fashion. 

{Index measurements for all member stars in K3 are listed in Table~\ref{tab:data}.}

\begin{figure*}
	\includegraphics[width=16cm]{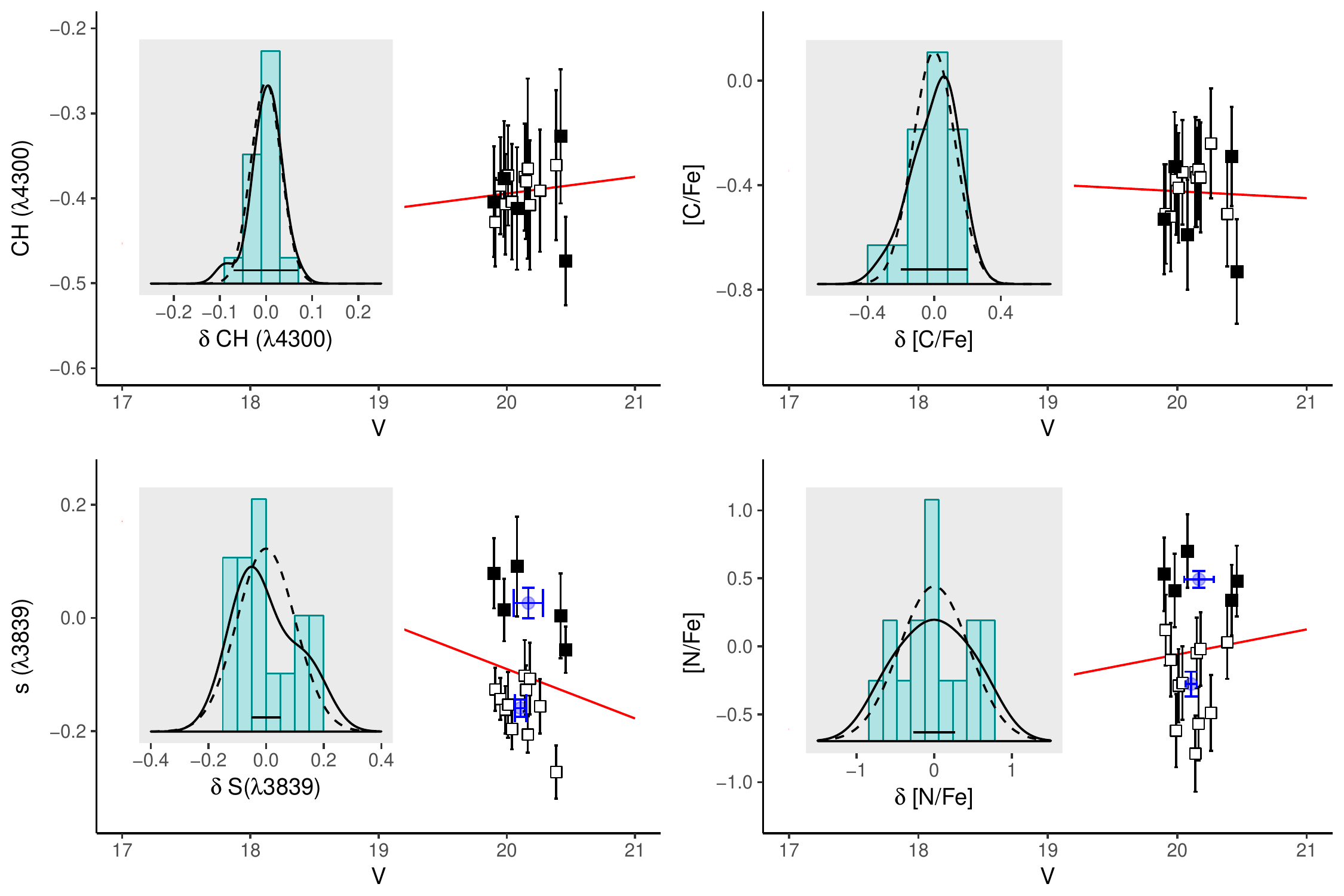}
	\caption{{\em From top to bottom, left to right:} The run of CN, CH, [C/Fe], and [N/Fe] is plotted against the apparent V band magnitude for all member stars. The red lines indicate the linear fit of those quantities vs. the visual magnitude. CN-strong and CN-weak stars are plotted as filled and empty symbols, respectively. The mean CN index and [N/Fe] abundance of the CN-strong and CN-weak groups and the associated standard errors are plotted in blue.
	The insets show the histograms and the associated kernel distributions (solid black line) of the residuals. The dashed line represents the Gaussian distribution that best fit the data. The mean errors associated with the measurements are also plotted at the base of each histogram.}

	\label{fig:cvnv}
\end{figure*}

\subsection{[C/Fe] and [N/Fe]}
\label{sec:candn}

In addition to using the CN and CH band strengths to investigate multiple populations in K3, we also used spectral synthesis to quantify [C/Fe] and [N/Fe] for the stars to look for evidence of an enriched population. We first evaluated effective temperatures ($T_{\rm{eff}}$) using the $(V-I)$ colour in the $T_{\rm{eff}}$-colour calibration by \citet{Alonso99}. For this calibration we adopted [Fe/H] = --1.08 from \citet{dacosta98} and $E(V-I) = 0.008$ from \citet{glatt08}. We then calculated the surface gravity for each star using the previously derived $T_{\rm{eff}}$ with a distance modulus of 18.8 \citep{glatt08}, the assumption of a mass of 0.95$\msun$ from BaSTI isochrones and the bolometric corrections from \citet{Alonso99}. Finally, a microturbulent velocity of  $v_{t}=2.0 \, kms^{-1}$ was assigned to all stars. 

Stellar abundances for [C/Fe] and [N/Fe] were calculated in the same way as in H17. We took line lists (both atomic and molecular) from the most recent Kurucz compilation from the website of F. Castelli\footnote{\url{http://wwwuser.oats.inaf.it/castelli/linelists.html}}. The grid of models used as starting points to determine model atmospheres was also taken from this website. The atmospheres were then calculated using the ATLAS9 code \citep{castelli04} with the previously quoted values for [Fe/H], $T_{\rm{eff}}$, v$_{\rm mic}$, $\log$(g) and a solar-scaled composition.

Model spectra with a range of abundances were generated with the SYNTHE code by Kurucz and subsequently fitted to our observations finding the best fit with a $\chi^{2}$ minimisation routine and thereby deriving [C/Fe] and [N/Fe] with solar abundances taken from \citet{Asplund09}. 
A fit was determined by minimising the observed-computed spectrum difference in a 100\AA~window centred on 4300\AA~for the CH G-band and 50\AA~window for the UV CN feature at 3883\AA. Running SYNTHE on quite a broad spectral range of about 200\AA~ to produce synthetic spectra allowed us to set a reasonable continuum level also by visual inspection and thus compute robust abundances.
It was necessary to use an iterative method as the C abundance must be known prior to determining that of N, as both contribute to the CN band.  

We adopted a constant oxygen abundance ([O/Fe]= +0.2 dex) throughout all computations. The derived C abundance is dependent on the O abundance and therefore so is the N abundance. To test the sensitivity of the C abundance to the adopted O abundance we varied the oxygen abundances and repeated the spectrum synthesis to determine the exact dependence for the coolest and warmest stars in our sample (with $T_{\rm{eff}} \sim $ 4600 and 5000, respectively. 
In these computations, we adopted [O/Fe]= 0.0 and +0.4 dex. We found that significantly large variations in the oxygen abundance slightly affect ($\delta$A(C)/
$\delta$[O/Fe] $\simeq$ 0.05 dex) the derived C abundance in cooler stars ($T_{\rm{eff}} \sim $ 4600-4800), while they are negligible (of the order of 0.02 dex or less) for hotter stars. The total error in the A(C) and A(N) abundance was computed by taking into account the two main sources of uncertainty: {\em (1)} the error in the adopted atmospheric parameters and {\em (2)} the error in the fitting procedure (continuum placement and random noise). Errors in the adopted $T_{\rm{eff}}$ translate to an uncertainty of $\delta$A(C)/ $\delta T_{\rm{eff}}$ $\simeq$ 0.05 - 0.10 dex and $\delta$A(N)/ $\delta T_{\rm{eff}}$ $\simeq$ 0.06 - 0.12 in the C and N abundances (similar for the hottest and coolest stars in our sample). The errors due to uncertainties on gravity are negligible (on the order of 0.06 dex or less) and those due to continuum placement are of the order of $\sim$0.15 dex.
To evaluate intrinsic errors on C and N measurement abundances the fitting procedure is repeated for a sample of 500 synthetic spectra where Poissonian noise has been introduced to reproduce the observed noise conditions.
The errors derived from the fitting procedure were then added in quadrature to the errors introduced by atmospheric parameters, resulting in an overall error of $\sim \pm$0.20 dex for the C abundances and $\sim \pm$0.27 dex for the N values.

 A combination of these errors was used as the final error estimation for each abundance measurement, as given in Table.~\ref{tab:data}.

\section{Results}
\label{sec:results}

In Fig.~\ref{fig:cvnv} we plot the CN and CH indexes along with the associated [C/Fe] and [N/Fe] abundance ratios for each of our member stars vs. their apparent V band magnitudes. The insets show the histograms of the residuals of the linear fit of those quantities vs. the visual magnitude.
In the case of both the CH index and [C/Fe] abundance ratio, the derived spread is very small and within the uncertainties. 
Thus, no significant carbon variation among target stars can be deduced from the data.

\begin{figure}
\centering
	\includegraphics[width=6.5cm]{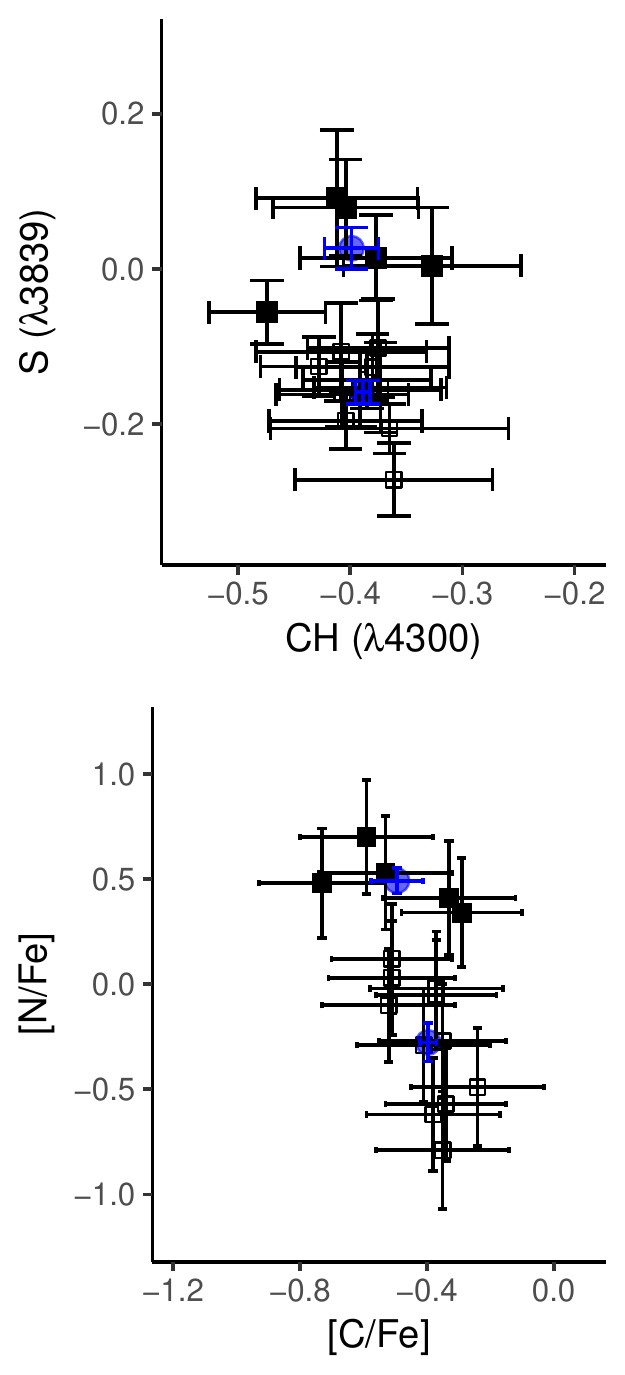}
		\caption{The top plot shows CH($\lambda$4300) against S($\lambda$3839) band strengths tracing nitrogen and carbon respectively for all member stars, while the bottom plot shows the same stars'  [C/Fe] against [N/Fe]. In both cases, there is clearly a larger spread in nitrogen than carbon. Mean abundances are also shown, along with their associated standard errors. Symbols are the same as Fig.~\ref{fig:cvnv}.}
	\label{fig:cnch}
\end{figure}

Conversely, a visual inspection of the bottom left hand panel of Fig~\ref{fig:cnch} reveals a statistically significant bimodality in the CN index over the whole sampled magnitude range. CN-strong and CN-weak stars are observed at the same magnitude, indicating that the observed star-to-star variation in CN must be intrinsic. Indeed, the observed bimodality in CN (as well as the elevated nitrogen abundances) cannot be explained by evolutionary mixing as they are fainter than the luminosity function bump \citep[$V_{BUMP} = 19.38 \pm 0.04$;][]{alves99}. Therefore, any mixing with evolution should have had little impact on our abundances for both carbon and nitrogen \citep[e.g.][]{gratton00}. An increase in CN band strength with increasing luminosity is possible, but such an observed trend relies only on a few targets at the faintest magnitude end. Using this plot we are able to clearly distinguish between the enriched (i.e. CN-strong stars, shown as filled symbols) and CN-weak (i.e. non-enriched, shown as empty symbols) component. The mean CN index and [N/Fe] abundance were then computed within each subpopulation. The difference in S($\lambda$3839) between CN-strong and CN-weak stars of comparable magnitude is $\sim$0.2 mag.

CN-strong stars tend also to have on average higher [N/Fe] abundance than CN-weak stars, with the mean [N/Fe] abundance of the two groups $\sim$ +0.5 and --0.3 dex respectively. In this case, because we observe a large spread rather than an obvious bimodality, computing the mean abundance of subpopulations split according to their CN index will tend to exaggerate the differences between the two populations. However, to assess the statistical significance of the internal nitrogen variation, we consider also the stars at the extremes of the distribution. Stars with the highest and lowest [N/Fe] abundances differ at more than  3$\sigma$ level. There is therefore additional evidence for internal variation in nitrogen in K3 stars also from nitrogen abundances despite the larger uncertainties associated to [N/Fe] measurements. The enhancement of the UV CN band which is observed in the CN-strong population with respect to CN-weak one 
cannot have been misinterpreted because of errors in the model-atmosphere abundance analysis, as shown in Fig.~\ref{fig:spectra}, where we plot the spectra of two stars with similar atmospheric parameters yet different strength of their CN absorption.

Both band strengths and abundances show the same trend: a negligible spread in carbon compared to a significant spread in nitrogen, which is greater than the errors. Again, the filled symbols indicate those that we would consider enriched in nitrogen. We conclude that the presence of such a CN-strong sub-population is a strong indication that multiple populations are present in Kron 3, as shown for Lindsay 1 in H17, as well as globular clusters in other galaxies \citep{kayser08}. The spread in nitrogen of up to 1 dex in Kron 3 is comparable to that of Lindsay 1, for which however, we were not able to detect a clear bimodality in the CN absorption bands.

Out of a sample of 16 stars, 5 have strong CN bands. This could suggest that the majority of stars belong to the non-enriched population, contrary to what is observed in Galactic GCs with present mass similar to that of K3. However, the spectroscopic sample presented here may be not representative of the entirety of the cluster, as the statistics are poor (16 stars) and the radial coverage is severely biased towards the outskirts of the cluster due to crowding in the central regions. From larger photometric samples, \citet{niederhofer17} find that the fraction of enriched stars ranges from 24\% to 45\% in three SMC clusters with masses and ages similar to those of K3 (ages $\sim$ 6-8 Gyr, M$\sim$ 10$^5$ $\msun)$.
\citet{dalessandro16} and \citet{florian} find that the non-enriched population account for more than 65\% of the total cluster mass also in NGC~121, the oldest cluster in the SMC at 11 Gyr, although due to differing methods, the fraction is not directly comparable to that of the HST UV Legacy Sample \citep{milone17}. Our result  agrees well with the number ratios given by \citet{florian,niederhofer17} and \citet{dalessandro16}, but more studies are needed to confirm whether the observed paucity of enriched stars in the SMC clusters with respect to the Milky Way clusters can be considered a general  characteristic of the SMC cluster population.

In Fig.~\ref{fig:cnch} we plot the run of the CN band vs. the CH absorption (top panel) and the [N/Fe] abundance ratios vs. [C/Fe] (bottom panel). The 
mean values along with the associated standard errors are also indicated. From this plot we note that CN-strong stars are not necessarily 
CH-weak (i.e. C and N are not anti-correlated, see also the top panel of Fig.~\ref{fig:cvnv}. This could possibly indicate that the nitrogen variations are not 
necessarily associated with carbon variations. 

Such a trend was already observed in L1 and also in Milky Way GCs by \citet{meszaros} from APOGEE data \citep{maj}. The former find that  a statistically significant anti-correlation  between [C/Fe] and [N/Fe] cannot be seen in any of the studied clusters, while a clear 
bimodality in CN can be observed for the most metal-rich ([Fe/H] $\leq$ --1.5 dex) clusters of the sample. This seems to be the case of both K3 and L1.
Our [C/Fe] abundance ratio determinations have large associated uncertainty of $\sim$ 0.20 dex. If the anti-correlations exist, they must span ranges smaller than our uncertainties in [C/Fe]. In general (within ancient GCs), nitrogen spans very large ranges (up to $\Delta$ [N/Fe] $\sim$ 2 dex), while [C/Fe] variations are usually smaller than $\Delta$ [C/Fe] $\sim$ 0.5 dex (see for example figures 10 and 11 in \citealp{cohen05}). Similar to \citet{meszaros}, we believe that the lack of any clear evidence of carbon variations is due to the spread being smaller than the precision of the instrument observing very faint stars. 


\section{Discussion and conclusions}
\label{sec:concs}

FORS2 spectroscopy of 35 RGB stars in the intermediate age massive cluster Kron 3 in the SMC have yielded 16 member stars after a conservative membership selection process.
We were able to derive CN and CH index measurements and C and N abundances for 16 stars. Five stars out of 16 show nitrogen enrichment. Figs.~\ref{fig:cvnv} and \ref{fig:cnch} indicate a significant spread in both S$\lambda$3839 and [N/Fe], larger than the errors, compared to a relatively constant CH$\lambda$4300 and [C/Fe]. As with our previous study of a similar intermediate age massive cluster in the SMC, Lindsay 1 (H17), we suggest that this spread is indicative of the presence of multiple populations. At 6.5 Gyr, Kron 3 is one of the youngest clusters to show these variations, which means that the still-unknown mechanism responsible for the onset of MPs operates until a redshift of at least 0.75, much later than the peak of globular cluster formation at redshift of $\approx 3$. This means the mechanism could still be working to create multiple populations in young massive clusters, so they can be considered analogues to GCs and used to constrain GC formation theories.

Multiple populations are usually identified using high resolution spectra yielding abundances of elements, most commonly [O/Fe] and [Na/Fe] \citep{carretta09}. Our method using low resolution spectra and band strengths is also routinely used to study MPs in ancient GCs \citep[e.g.][]{kayser08,martell08,pancino10}. This method has some advantages over high resolution spectroscopy, with a simpler and quicker method for obtaining results and the ability to study the presence of MPs in clusters at greater distances that are too faint to be studied at high resolution with a reasonable observing time. To ensure that our results are reliable, we have used strict membership criteria as detailed in \S~\ref{sec:members} and ensured our abundances and band strengths are not affected by effects such as evolutionary mixing. 

This method has already been validated in the past for the old, low mass stars: the MW GCs that show both photometric evidence of multiple populations and have spectroscopically confirmed presence of MPs through Na-O and similar light element anti-correlations have been shown to present variations in CN index and/or [N/Fe] spread \citep[e.g.][]{pancino10}. Furthermore, this method was used to detect MPs in the intermediate-age cluster L1, which has younger and thus more massive RGB stars than typical ancient GCs. The spread in nitrogen reported by H17 was confirmed to correspond to different populations via HST photometry \citep{niederhofer17}. This strengthens the reliability of this method and our conclusion that Kron 3 hosts multiple populations.

\section*{Acknowledgements}
C.L. thanks the Swiss National
Science Foundation for supporting this research through the
Ambizione grant number PZ00P2 168065.

E.K.G. gratefully acknowledges support from the German Research
Foundation (DFG) via Sonderforschungsbereich SFB 881 (``The Milky
Way System'', subproject A8).

\bibliographystyle{mnras}
\bibliography{biblio}

\bsp	
\label{lastpage}

\clearpage\pagestyle{empty}

\addtolength{\textheight}{0.65in}

\begin{landscape}
	\begin{table}
		\centering
		Stellar properties
		\begin{tabular}{rcccccccccrcrccccrccrc}
		
		\hline
		\hline
Star   & R.A.  & Dec 	  &     V    &    T$_{{\rm eff}}$     &  eT$_{{\rm eff}}$ &  log(g)   &elog(g)&    [C/Fe]   & e[C/Fe]   &   [N/Fe]	& e[N/Fe]    &   CN    &   eCN & CH  &   eCH &     Ca(HK)   &   eCa  &  Fe	    &    eFe &   RV	 & Notes \\
	 & (Degrees) & (Degrees)	&       (mag)      &    (K)      & 	(K)    	&    (dex)  &(dex) & (dex)     &	(dex)	     & (dex)  & (dex)       & 	       (mag)                  & 	(mag)	              & 	(mag)	               & 	(mag)	                &   (mag)                     &  (mag)                        &       (mag)    &   (mag)        &  (km/s) &      \\
			  \hline
 			  
  225   &    6.135502  &  --72.82457    &   19.99   &   4837    &     90  &   2.4  &  0.2   &    --0.38  &   0.21  & --0.62  &  0.27  &  --0.162  &   0.042  &   --0.407   &   0.059  &    25.773  &   6.265   &     --0.165	 &    0.03   &   123.0   &   1P       \\ 
  419   &    6.298478 &   --72.85623   &     20.16  &   4607    &     78  &   2.4  &  0.2   &    --0.34  &   0.19  & --0.57  &  0.27  &  --0.206  &   0.032  &   --0.365   &   0.106  &    17.181  &  11.652   & 	--0.153   &    0.02  &   122.1   &   1P       \\ 
  599   &    6.225902 &   --72.84796  &      20.39  &   5000    &     98  &   2.7  &  0.2   &    --0.51  &   0.20  &  0.03  &  0.27  &   --0.272  &   0.047  &   --0.361   &   0.088  &    23.915  &   7.979   & 	--0.172   &    0.02  &   115.3   &   1P       \\ 
  1517  &   6.292014  &  --72.81255   &     20.42   &   4907    &     93  &   2.7  &  0.2   &   --0.29  &   0.19  &  0.34  &  0.26  &    0.004  &   0.075  &   --0.327   &   0.079  &    20.342  &    9.493   &    --0.191	 &    0.03   &   142.1   &   2P       \\ 
  1903  &   6.267869  &  --72.80966   &     20.46   &   4980    &     97  &   2.7  &  0.2   &   --0.73  &   0.20  &  0.48  &  0.26  &   --0.056  &   0.041  &   --0.474   &   0.052  &    20.352  &   9.258   &     --0.166	 &    0.03   &   141.9   &   2P       \\ 
  2349  &   6.115903  &  --72.80645    &    20.15   &   4791    &     87  &   2.5  &  0.2   &   --0.37  &   0.19  & --0.05  &  0.26  &  --0.127  &   0.043  &   --0.380    &   0.068  &    25.076  &   8.199   & 	--0.176   &    0.03   &   140.3   &    1P      \\ 
  2680  &   6.156927  &  --72.80423    &    20.14   &   4814    &     88  &   2.5  &  0.2   &   --0.35  &   0.21  & --0.79  &  0.28  &  --0.102  &   0.063  &   --0.375   &   0.063  &    22.207  &   9.929   & 	--0.176   &    0.03   &   141.6   &    1P      \\ 
  3130  &   6.231442  &  --72.80103   &     19.91   &   4883    &     92  &   2.4  &  0.2   &   --0.51  &   0.19  &  0.12  &  0.26  &   --0.126  &   0.038  &   --0.428   &   0.052  &    17.637  &   5.127   & 	--0.191   &    0.03   &   131.3   &    1P      \\ 
  3546  &   6.219912  &  --72.79807   &     20.08   &   4747    &     85  &   2.4  &  0.2   &   --0.59  &   0.21  &  0.70  &  0.27  &    0.091  &   0.088  &   --0.412   &   0.072  &    21.487  &   6.405   & 	--0.153   &    0.03   &   130.2   &    2P      \\ 
  3910  &   6.219479  &  --72.79566   &   19.98   &   4814    &     88  &   2.4  &  0.2   &   --0.33  &   0.21  &  0.41  &  0.27  &      0.014  &   0.055  &   --0.377   &   0.068  &    23.889  &   6.953   & 	--0.176   &    0.03   &   145.4   &    2P      \\ 
  5174  &   6.245548  &  --72.78617   &   19.95   &   4791    &     87  &   2.4  &  0.2   &   --0.52  &   0.21  & --0.10  &  0.27  &   --0.143  &   0.037  &   --0.385   &   0.057  &    22.241  &   5.330	 & 	--0.174   &    0.03   &   141.2   &    1P      \\ 
  6130  &   6.175249  &  --72.77926   &   20.18   &   4837    &     90  &   2.5  &  0.2   &   --0.37  &   0.21  & --0.02  &  0.27  &   --0.107  &   0.063  &   --0.408   &   0.076  &    22.867  &   8.077   & 	--0.194   &    0.03   &   151.3   &    1P      \\ 
  6463  &   6.248931  &  --72.77658   &   20.26   &   4791    &     87  &   2.5  &  0.2   &   --0.24  &   0.21  & --0.49  &  0.28  &   --0.156  &   0.048  &   --0.391   &   0.072  &    20.158  &   8.311   & 	--0.173   &    0.03   &   148.2   &    1P      \\ 
  6862  &   6.206618  &  --72.77339   &   19.90   &   4769    &     86  &   2.3  &  0.2   &   --0.53  &   0.21  &  0.53  &  0.27  &     0.079  &   0.062  &   --0.404   &   0.065  &    24.803  &   8.432   & 	--0.171   &    0.03   &   159.9   &    2P      \\ 
  7293  &   6.236802  &  --72.76972   &   20.01   &   4683    &     82  &   2.3  &  0.2   &   --0.41  &   0.21  & --0.29  &  0.27  &  --0.153  &   0.058  &   --0.373   &   0.059  &    27.081  &   9.327   & 	--0.177   &    0.03   &   141.2   &    1P      \\ 
  7963  &   6.302150  &  --72.76238   &   20.04   &   4883    &     92  &   2.5  &  0.2   &   --0.35  &   0.20  & --0.27  &  0.27  &  --0.196  &   0.036  &   --0.404   &   0.068  &    20.208  &   7.894   & 	--0.206   &    0.03   &   148.3   &    1P      \\ 
  220   &    6.271843  &  --72.86605  &   21.18  &    \ldots    &    \ldots   &    \ldots  &   \ldots   &     \ldots   &   \ldots   &   \ldots  &	\ldots  &   \ldots	&    \ldots   & \ldots	&    \ldots   &	 \ldots	&    \ldots    &	     \ldots    &	 \ldots	&  145.8    &  member	 \\ 
  285   &    6.260201  &  --72.86331  &   20.76  &    \ldots    &    \ldots   &    \ldots  &   \ldots   &     \ldots   &   \ldots   &   \ldots  &	\ldots  &   \ldots	&    \ldots   &	 \ldots	&    \ldots   &	 \ldots	&    \ldots    &	     \ldots    &	 \ldots	&  106.1    &  member	 \\ 
  791   &    6.207983  &  --72.83820  &   19.72  &    \ldots    &    \ldots   &    \ldots  &   \ldots   &     \ldots   &   \ldots   &   \ldots  &	\ldots  &   \ldots	&    \ldots   &	 \ldots	&    \ldots   &	 \ldots	&    \ldots    &	     \ldots    &	 \ldots	&  107.4    &  member	 \\ 

  129   &    6.112807 &   --72.87157  &   20.95  &    \ldots    &    \ldots   &    \ldots  &   \ldots   &     \ldots   &   \ldots   &   \ldots  &	\ldots  &   \ldots	&    \ldots   &	 \ldots	&    \ldots   &	 \ldots	&    \ldots    &	     \ldots    &	 \ldots	&  118.9    &  field	 \\ 
  176   &    6.293897 &   --72.86896  &   20.57  &    \ldots    &    \ldots   &    \ldots  &   \ldots   &     \ldots   &   \ldots   &   \ldots  &	\ldots  &   \ldots	&    \ldots   &	 \ldots	&    \ldots   &	 \ldots	&    \ldots    &	     \ldots    &	 \ldots	&  130.9    &  field	 \\ 
  286   &    6.143601 &   --72.87489  &   19.57   &    \ldots   &    \ldots   &    \ldots  &   \ldots   &     \ldots   &   \ldots   &   \ldots  &	\ldots  &   \ldots	&    \ldots   & \ldots	&    \ldots   &	 \ldots	&    \ldots    &	     \ldots    &	 \ldots	&  217.5    &  field	 \\ 
  329   &    6.066787 &   --72.86083  &   20.76  &    \ldots    &    \ldots   &    \ldots  &   \ldots   &     \ldots   &   \ldots   &   \ldots  &	\ldots  &   \ldots	&    \ldots   & \ldots	&    \ldots   &	 \ldots	&    \ldots    &	     \ldots    &	 \ldots	&  89.5     &  field	 \\ 
  514   &    6.260530  &  --72.82125   &   20.09   &    \ldots    &    \ldots   &    \ldots  &   \ldots   &     \ldots   &   \ldots   &   \ldots  &	\ldots  &   \ldots	&    \ldots   &	 \ldots	&    \ldots   &	 \ldots	&    \ldots    &	     \ldots    &	 \ldots	&  174.8    &  field	 \\ 
  651   &    6.227692 &   --72.84573    &   19.91  &    \ldots    &    \ldots   &    \ldots  &   \ldots   &     \ldots   &   \ldots   &   \ldots  &	\ldots  &   \ldots	&    \ldots   &	 \ldots	&    \ldots   &	 \ldots	&    \ldots    &	     \ldots    &	 \ldots	&  90.6     &  field	 \\ 
  715   &    6.247569 &   --72.84168    &   21.12   &    \ldots    &    \ldots   &    \ldots  &   \ldots   &     \ldots   &   \ldots   &   \ldots  &	\ldots  &   \ldots	&    \ldots   &	 \ldots	&    \ldots   &	 \ldots	&    \ldots    &	     \ldots    &	 \ldots	&  95.5     &  field	 \\ 
  825   &    6.213307  &  --72.81841    &   20.57   &    \ldots    &    \ldots   &    \ldots  &   \ldots   &     \ldots   &   \ldots   &   \ldots  &	\ldots  &   \ldots	&    \ldots   &	 \ldots	&    \ldots   &	 \ldots	&    \ldots    &	     \ldots    &	 \ldots	&  112.0    &  field	 \\ 
  842   &    6.241302  &  --72.83541    &   20.46  &    \ldots    &    \ldots   &    \ldots  &   \ldots   &     \ldots   &   \ldots   &   \ldots  &	\ldots  &   \ldots	&    \ldots   &	 \ldots	&    \ldots   &	 \ldots	&    \ldots    &	     \ldots    &	 \ldots	&  95.0     &  field	 \\ 
  905   &    6.257216 &    --72.83270   &   20.73  &    \ldots    &    \ldots   &    \ldots  &   \ldots   &     \ldots   &   \ldots   &   \ldots  &	\ldots  &   \ldots	&    \ldots   &	 \ldots	&    \ldots   &	 \ldots	&    \ldots    &	     \ldots    &	 \ldots	&  152.2    &  field	 \\ 
  1201  &   6.194623  &   --72.81501   &   20.83   &    \ldots    &    \ldots   &    \ldots  &   \ldots   &     \ldots   &   \ldots   &   \ldots  &	\ldots  &   \ldots	&    \ldots   &	 \ldots	&    \ldots   &	 \ldots	&    \ldots    &	     \ldots    &	 \ldots	&  94.3     &  field	 \\ 
  4281  &   6.236492  &   --72.79306   &   20.66   &    \ldots    &    \ldots   &    \ldots  &   \ldots   &     \ldots   &   \ldots   &   \ldots  &	\ldots  &   \ldots	&    \ldots   &	 \ldots	&    \ldots   &	 \ldots	&    \ldots    &	     \ldots    &	 \ldots	&  94.8     &  field	 \\ 
  4542  &   6.272315  &   --72.79084   &   20.11   &    \ldots    &    \ldots   &    \ldots  &   \ldots   &     \ldots   &   \ldots   &   \ldots  &	\ldots  &   \ldots	&    \ldots   &	 \ldots	&    \ldots   &	 \ldots	&    \ldots    &	     \ldots    &	 \ldots	&  176.8    &  field	 \\ 
  4873  &   6.270678  &   --72.78851   &   20.92   &    \ldots    &    \ldots   &    \ldots  &   \ldots   &     \ldots   &   \ldots   &   \ldots  &	\ldots  &   \ldots	&    \ldots   &	 \ldots	&    \ldots   &	 \ldots	&    \ldots    &	     \ldots    &	 \ldots	&  138.4    &  field	 \\ 
  5697  &   6.220817  &   --72.78246   &   20.28   &    \ldots    &    \ldots   &    \ldots  &   \ldots   &     \ldots   &   \ldots   &   \ldots  &	\ldots  &   \ldots	&    \ldots   &	 \ldots	&    \ldots   &	 \ldots	&    \ldots    &	     \ldots    &	 \ldots	&  213.6    &  field	 \\ 
  7604  &   6.231820   &  --72.76645   &   20.23   &    \ldots    &    \ldots   &    \ldots  &   \ldots   &     \ldots   &   \ldots   &   \ldots  &	\ldots  &   \ldots	&    \ldots   &	 \ldots	&    \ldots   &	 \ldots	&    \ldots    &	     \ldots    &	 \ldots	&  166.7    &  field	 \\ 
 &     &      &      &      &       &   &     &    &     &   &    &    &     &      &     &      &    &     	 &      &      &         \\ 

\hline

		\end{tabular}
		\caption{Table of stellar properties for all 35 stars in our programme. Stars that were found to be non-members using the criteria specified in \S~\ref{sec:members} are indicated. The faintest stars 220, 285, and 791 are excluded from our analysis due either to their low S/N and the presence of artifacts/defects within the molecular spectra.}
		\label{tab:data}
	\end{table}
\end{landscape}

\pagestyle{plain}











\end{document}